\documentclass[11pt,a4paper]{article}
\usepackage[]{amsmath,amssymb}
\usepackage{graphics,epsfig}
\newcommand{\be}{\begin{equation}}
\newcommand{\ee}{\end{equation}}
\newcommand{\beq}{\begin{equation}}
\newcommand{\eeq}{\end{equation}}
\newcommand{\bea}{\begin{eqnarray}}
\newcommand{\eea}{\end{eqnarray}}
\newcommand{\nn}{\nonumber}
\def\be{\begin{equation}}
\def\ee{\end{equation}}
\def\ba{\begin{eqnarray}}
\def\ea{\end{eqnarray}}

\begin{document}
\baselineskip=15.5pt \pagestyle{plain} \setcounter{page}{1}

\begin{titlepage}
\begin{centering}

{\Large {\bf Hydrodynamic modes of a holographic $p-$wave
superfluid}}

\vspace{.3in}

Ra\'ul E. Arias $^{\dag}$\footnote{rarias@cab.cnea.gov.ar}  and Ignacio Salazar Landea $^{\ddag}$\,\footnote{peznacho@gmail.com} \\
\vspace{.2 in}
$^{\dag}${\it Centro At\'omico Bariloche\\ 8400-S.C. de Bariloche\\
R\'{\i}o Negro, Argentina}\\
~

$^{\ddag}${\it IFLP-CONICET and Departamento de F\'{\i}sica\\ Facultad de Ciencias Exactas, Universidad Nacional de La Plata\\
CC 67, 1900,  La Plata, Argentina}\\
~

\vspace{.4in}

{\bf Abstract} \\

\end{centering}

In this work we analyze the hydrodynamics of a $p-$ wave superfluid
on its strongly coupled regime by considering its holographic
description. We obtain the poles of the retarded Green function
through the computation of the quasi-normal modes of the dual AdS
black hole background finding diffusive, pseudo-diffusive and sound modes. For
the sound modes we compute the speed of sound and its attenuation as
function of the temperature. For the diffusive and pseudo-diffusive
modes we find that they acquire a non-zero real part at certain
finite momentum.
\end{titlepage}

\tableofcontents

\section{Introduction}

The gauge/gravity correspondence \cite{Malda, gkpw} provides a
useful tool to study quantum field theories in the strongly coupled
regime. In the last years, this was applied to describe the behavior
of condensed matter systems (see \cite{Hartnoll, McG, Musso:2014efa}
for nice reviews). In the present work we will study the
hydrodynamic modes of a $p$-wave superconductor using holography. In \cite{Hoyos} the authors for a studied this system
in QFT.

The order parameter for this class of superconductors is a vector
field. Its gravity dual was proposed in the probe limit in
\cite{Gubser} and the backreaction on the metric was analyzed in
\cite{Basu:2009vv, Arias, Ammon:2009xh}. Holographically the $p-$
wave superconductor \cite{Gubser} can be modeled with an
asymptotically AdS black hole with a $SU(2)$ gauge field living on
it. The boundary behavior of this gauge field gives the chemical
potential and the order parameter of the superconductor. In
particular, the chemical potential breaks the $SU(2)$ gauge group to
a $U(1)$ subgroup and the condensate spontaneously breaks this
$U(1)$ subgroup and rotational symmetry. In this work we are going
to take into account the probe limit approximation in which the
gauge field doesn't deform the geometry of the space-time. Roughly
speaking this approximation means that we have a small amount of
charged (under the gauge group) degrees of freedom compared to the
total number of degrees of freedom of the QFT.

The hydrodynamic theory studies the conservation equations for whatever symmetry the system has.
In particular the conservation of the stress energy tensor and, if the system has some global symmetry,
 the conservation of the associated Noether current. The ansatz that solves these systems are called constitutive equations
  and because hydrodynamics is an effective theory they are functions of the derivatives of the temperature,
   the local velocity and the chemical potential. From those equations we can compute some transport properties
   like the diffusion constant, the first and second sound velocity, the charge conductivity,
   the charge density and the shear and bulk viscosities.
   Linear response theory allow us to know the dispersion relation for the hydrodynamics modes through
    the poles of the retarded Green function
   (see \cite{Kovtun} for a nice review on relativistic hydrodynamics).
   Since we will work in the probe limit, we will only have
   information about the fluctuations of the modes associated to the
  Noether current. A more complete study of the system with
  backreaction could enlighten us about the stress energy tensor
  fluctuations.

As was shown in \cite{Son} the retarded Green function can be
obtained holographically through the study of the quasinormal modes
(QNM's) of the dual gravity background (see \cite{qnm} for reviews
on the computation of QNM's in black hole geometries). Previous
works on holographic hydrodynamics compute the hydrodynamic modes of
$\cal{N}$ $=4$ SYM \cite{Policastro}. The aim of the present work is
to analyze the hydrodynamic modes of the $p-$wave superconductor
through the study of the quasinormal modes spectrum of its
holographic dual. In other words, we are going to study the poles of
the current-current two point functions $\langle j_i^a\,
j_i^a\rangle$ in the probe limit approximation. Here, the current
$j_x^1$ is the order parameter of the $p-$wave superconductor.
Moreover we are interested in the computation of the second sound of
the superconductor. In the two fluid model of superconductivity the
system is divided in two components, the normal fluid that contains
elementary excitations like the phonon and roton and the superfluid
that consists on the condensate. This model predicts the existence
of a second sound mode which is the de-phased collective motion of
the two fluids. We are going to use holography to compute this
second sound in a $p-$wave superconductor (see  \cite{Herzog, Amado}
for previous work on this direction). Moreover we are going to
obtain diffusive and pseudo-diffusive modes.

In \cite{Gauntlett} the authors did a Gaussian quench in the order
parameter of the gravity dual of a $s-$wave superfluid \cite{HHH}.
They showed that a universal late time behavior of the system is
described by the poles of the retarded Green function that lie the
closest to the real axis. Furthermore, it is discussed in
\cite{Amado:2013xya} that the dynamical phase diagram can be
extracted from the information of the QNM at equilibrium. Then, this
is another application of our study of the quasinormal modes. They
will describe the late time behavior of the quenched $p-$wave
superconductor.

The program is the following. In Section \ref{model} we are going to
review the gravity dual to the $p-$wave superconductor in the probe
limit proposed on \cite{Gubser}. In Section \ref{qnm} we study the
quasinormal spectrum and describe the second sound, the diffusive
and pseudo-diffusive modes. Finally, the results are summarized in
Section \ref{concl}.

\section{Holographic $p-$wave superconductor}
\label{model}

 In this section we are going to review the gravity
dual found in \cite{Gubser} for
a holographic $p-$wave superfluid.% and we study the gauge field fluctuations
%that we will analyze in order to get with the QNM spectrum.

The order parameter in conventional superconductors comes from
electron pairs that couple forming Cooper pairs. The state
describing this pair must be symmetric and then only certain
combinations are allowed. The $p-$wave superconductor is the name
for those systems in which the relative orbital angular momentum
between the electrons forming the Cooper pair is $l=1$. The
superfluid $^3He-A$ is a real world example of a system with this
order parameter.

We will work in the simplest set up and consider $SU(2)$ as the
gauge group. We will consider the system charged under the $U(1)$
inside $SU(2)$. This will explicitly break $SU(2)$ down to $U(1)$.
The $p$-wave superconductor ansatz, breaks the remaining $U(1)$
symmetry and the $SO(2)$ symmetry associated to spatial rotations in
the bidimensional boundary theory. The gravity solution that
describes the strong coupling dynamics is as follows: a charged
superconducting layer develops outside the horizon due to the
interplay between the electric repulsion %(with the charged black
%hole) guarda q en el probe limit el agujero negro no esta cargau
 and the gravitational potential of the asymptotically AdS
geometry. At high enough temperatures there is no hair outside the
black hole and the solution is just a charged AdS black hole. Below
a critical temperature $T_c$ a non-trivial gauge field with
non-vanishing chemical potential on the boundary of the geometry and
a sourceless non-vanishing condensate in the bulk appears,
 breaking the remaining $U(1)$ gauge symmetry. An alternative
 formulation for a $p-$wave superfluid can be found in
 \cite{Cai:2013aca}.

\subsection{The model}

%One advantage in the gravity dual to the $p-$wave superconductor
%over the dual to a $s-$wave is that in the former the bulk
%Lagrangian is fixed by symmetry and in the second one must add by
%hand a potential for the scalar field (remember that in a $s-$wave
%superconductor the order parameter is a scalar field \cite{HHH}).
%Another advantage is that this model could be easily embedded in
%string theory. Sure? porque aparecieron modelos nuevos de p waves y tb hay HHH en string theory

\noindent The Einstein-Yang-Mills action fixed by the gauge symmetry
reads \be S=\frac{1}{2\kappa^2}\int
d^4x\left(R-\frac14(F_{\mu\nu}^a)^2+\frac{6}{L^2}\right), \ee where
$\kappa$ is the gravitational constant in four dimensions and
$F_{\mu\nu}$ is the field strength of an $SU(2)$ gauge field \be
F_{\mu\nu}^a=\partial_\mu A_\nu^a-\partial_\nu
A_\mu^a+g_{_{YM}}\epsilon^{abc}A_\mu^bA_\nu^c\,. \ee Here the index
$a$ runs through the three $SU(2)$ generators. We are going to work
in the probe limit, that is large $\frac{g_{_{YM}}}{\kappa}$. By
scaling the gauge field as $\tilde A= \frac{A}{g_{_{YM}}}$ we see
that the large $g_{_{YM}}$ limit corresponds to the probe
(non-backreacting) limit of the gauge field. Roughly speaking one can think
that $\frac{1}{g_{_{YM}}^2}$ counts the degrees of freedom of the
dual field theory that are charged under the $SU(2)$ gauge group and
$\frac{1}{\kappa_{(4)}^2}$ counts the total number of degrees of
freedom. Then the probe limit means that we have a small number of
charged degrees of freedom with respect to the total number.
Moreover, taking this limit allow us to decouple the metric
fluctuations from the gauge field fluctuations. This means that we
are going to study the retarded Green functions for current-current
expectation values, $G^R_{jj}$, and we are not going to catch up the
transport coefficients that comes from the stress-energy tensor
conservation. In this limit, the proposed background geometry reads
\be
ds^2=\frac{r^2}{L^2}\left(-f(r)dt^2+dx^2+dy^2\right)+\frac{L^2}{r^2
f(r)}dr^2, \ee with $f(r)=1-\frac{r_h^3}{r^3}$ and $r_h$ standing
for the location of the black hole horizon. From now on we set the
scale $r_h=1$. The asymptotically AdS boundary is located at
$r=\infty$ and the temperature of the horizon is $T_h=\frac{3}{4\pi
L^2}$. The ansatz for the gauge field has the form \be
A=\phi(r)\tau^3dt+w(r)\tau^1dx.\label{gauge} \ee Here
$A=A_\mu^a\tau^adx^\mu$ with $\tau^a=\frac{\sigma^a}{2i}$ and
$\sigma^a$ the usual Pauli matrices. These $SU(2)$ generators
satisfy the standard algebra $[\tau^a,\tau^b]=\epsilon^{abc}\tau^c$.
A solution developing $w(r)\neq0$ in the gauge field ansatz breaks
the remaining $U(1)$ gauge symmetry associated with rotations around
$\tau^3$ (usually called $U(1)_3$). We are looking for solutions
that break this $U(1)_3$ symmetry spontaneously and we are going to
achieve this imposing a non trivial regular profile for the gauge
field.

 We will redefine our fields $\tilde\phi(r)=g_{_{YM}}L^2\phi(r),
\tilde w(r)=g_{_{YM}}L^2 w(r)$ in order to simplify the equations. This is equivalent to set $ g_{_{YM}}=L=1 $.
 The Maxwell equations $D_\mu F^{\mu\nu}=0$ on this geometry read
\bea
\tilde\phi''(r)+\frac{2}{r}\tilde\phi'(r)-\frac{\tilde w(r)^2\tilde\phi(r)}{r^4f(r)}=0\,,\\
\tilde w''(r)+\left(\frac{f'(r)}{f(r)}+\frac2r\right)\tilde
w'(r)+\frac{\tilde w(r)\tilde\phi(r)^2}{r^4f(r)^2}=0\,.\label{eomp}
\eea
%\bea
%\phi''(r)+\frac{2}{r}\phi'(r)-\frac{g_{_{YM}}^2L^4w(r)^2\phi(r)}{r^4f(r)}=0,\\
%w''(r)+\left(\frac{f'(r)}{f(r)}+\frac2r\right)w'(r)+\frac{g_{_{YM}}^2L^4w(r)\phi(r)^2}{r^4f(r)^2}=0.\label{eomp}
%\eea

The system \eqref{eomp} has the following behavior near the
horizon
\bea
\tilde\phi(r)\approx\phi_1^h(r-1), ~~~~~~~~~~~~~r\rightarrow 1\,,\nn\\
\tilde w(r)\approx w_0^h+w_2^h(r-1)^2,~~~~~~~~~r\rightarrow 1\,.
\eea On the other hand the near the boundary behavior for these
equations reads \bea
\tilde\phi(r)&=&\mu+\frac{\rho}{r},~~~~~ r\rightarrow \infty\,,\nn\\
\tilde w(r)&=&\frac{<j_x^1>}{r},~~~ r\rightarrow \infty\,. \eea
where $\mu$ is the chemical potential and $\rho$ the charge density
of the dual field theory. The expectation value of the current
$<j_x^1>$ is the order parameter of the superfluid phase.

We will express our results in the grand canonical ensemble, i.e. at
fixed chemical potential. Then the physical temperature of the
system will be determined by the following re-scaled dimensionless
magnitude $T=T_h/\mu$.

Note that in order to have a gravity dual of a spontaneously broken
symmetry we need to have the leading term $w_0^b\, r^0=0$ in the
boundary behavior of $\tilde w(r)$ because this is, according to the
AdS/CFT dictionary, the source for $<j_x^1>$. The solution is found
using a shooting technique %in the following form: suppose we have
%$w_0^b(\phi_1^h,w_0^h)\neq0$,
and the desired solution is obtained for values of the horizon
coefficients $\phi_1^h,w_0^h$ that satisfy
$w_0^b(\phi_1^h,w_0^h)=0$.

\begin{figure}[t!]
\centering
\includegraphics[width=230pt]{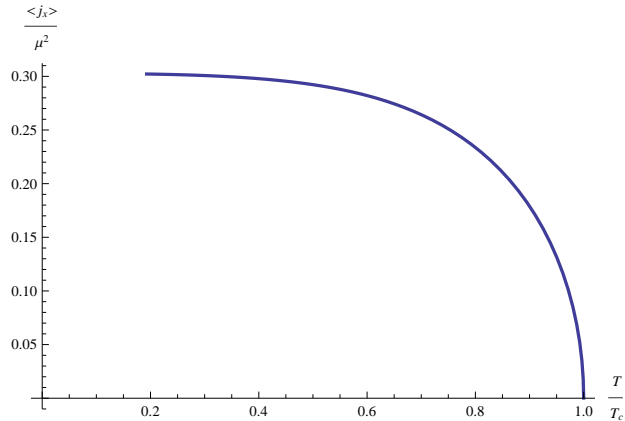}
\caption{Re-scaled order parameter $<j_x^1>/\mu^2$ as a function of
the temperature showing a second order phase transition for low
enough temperatures.} \label{condensate}\end{figure}

Exploring the space of parameters we find that a solution with
$w(r)\neq 0$ only exist for low enough temperatures, which
translates into a phase transition from a normal to a broken phase
characterized by the order parameter $<j_x^1>$, as shown in Figure
\ref{condensate}.

The free energy for this solution was computed in \cite{Gubser},
certifying that the broken phase is indeed thermodynamically
preferred. When going to lower temperatures than those shown in
Figure \ref{condensate} the probe limit breaks down and backreaction
on the metric must be considered \cite{Basu:2009vv, Arias,
Ammon:2009xh}. For analytical computations in this system see
\cite{Gangopadhyay:2012gx}

In what follows of this section we are going to consider
fluctuations of the gauge field.

\subsection{Gauge field fluctuations}

In order to study the QNM spectrum of this background we are going to consider the following fluctuations of the $SU(2)$ gauge field
\bea
{\cal A}&=&\left(\phi(r)\tau^3+\lambda \sum_{i=1}^3\delta A_t^i(t,x,y,r)\tau^i\right)dt+\left(w(r)\tau^1+\lambda\sum_{i=1}^3\delta A_x^i(t,x,y,r)\tau^i\right)dx\nn\\
&&+\lambda\sum_{i=1}^3\delta A_y^i(t,x,y,r)\tau^i dy\,, \eea where
we fix the gauge $a_r^i=0$ and $\lambda$ is the expansion
coefficient. Since we are interested in the QNM spectrum we will
work at linear order in $\lambda$.

To have simpler equations of motion for this fluctuations we can
divide them in different modes. In particular we are going to study
modes that propagate in a longitudinal or transverse direction with
respect to the direction of the condensate\footnote{Note that both
sectors should coincide in the zero momentum limit and differences
should arise only when considering $k\neq0$.}.

~

{\bf Longitudinal modes}

\be \delta A_{t,x,y}^i(t,x,y,r)=a_{t,x,y}^i(r)e^{-i \omega t+ i k_x
x}\,,\label{longitudinal} \ee

~

{\bf Transverse modes}

\be \delta A_{t,x,y}^i(t,x,y,r)=a_{t,x,y}^i(r)e^{-i \omega t+ i k_y
y}\,,\label{transversal} \ee

 ~

Analyzing the modes in an arbitrary direction in the fashion of
\cite{Amado:2013aea, Jokela:2014wsa} would be interesting but would
require a bigger computational power.

In the next section we are going to study the solution to the
equations of motion for the fluctuations $a_{t,x,y}^i(r)$ imposing
ingoing boundary conditions at the horizon of the black
hole\footnote{This ensures that we are dealing with retarded Green
functions. On the other hand, outgoingness at the horizon
corresponds to the holographic computation of the advanced Green
functions.}
%()
of the form \bea a_{x,\,y}^i&\approx&
a_{x_h,\,y_h}^i(r-1)^{-\frac{i\omega}{3}}+\cdots\,,
\\
a_{t}^i&\approx&a_{t_h}^i(r-1)^{-\frac{i\omega}{3}+1}+\cdots\,. \eea
At the boundary we are going to impose that the leading behavior
vanishes. This requirement will quantize the frequency, giving us
the quasinormal spectrum of the system.

These quasinormal frequencies are the poles of the retarded Green
function in the dual field theory. Furthermore, we are going to
study the small momentum regime of these modes, which will give us
the hydrodynamic regime of the $p-$wave superfluid.

Generically the equations of motion for the perturbations will
appear coupled in sectors. To deal with this issue we will use the
determinant method developed in \cite{Amado} in order to solve the
system. The desired quasinormal frequencies $\omega(k_i)$ are
obtained from the roots of the determinant of the matrix formed
with solutions to the equations evaluated at the boundary. In order
to get an squared matrix, we will need an independent solution for
each coupled equation.

Typically the number of linearized equations of motion for the
fluctuations is greater than the number of free horizon parameters.
This is due to constraint equations that relate some near horizon
parameters. When this happens we will take advantage of the
existence of (algebraic) pure gauge solutions in order to get as
many independent solutions as equations of motion.

\section{Quasinormal modes}\label{qnm}

In this section we are going to find the spectrum of exitations of
the $p-$wave
superconductor in the hydrodynamic limit. %We want to mention first
%that these modes should be understood as massless modes in the sense
%that $\mathrm{lim}_{k\rightarrow 0}\omega(k)=0$. Not all of them my friend
The analysis of these hydrodynamics modes can be done studying the
QNM frequencies of the gravity dual. Here we are going to study the
QNM spectrum of the geometry reviewed in Section \ref{model} and we
will numerically solve the equations of motion for the fluctuations
\eqref{longitudinal} and \eqref{transversal}. In particular we are
going to study its solutions in the unbroken (normal) and broken
(superfluid) phase. As a first step we will study the normal phase
which is going to give us information about the number of modes that
we have. In a second step we will follow this modes in the
superconducting phase and study their behavior.

\subsection{Longitudinal modes}

In this subsection we are going to study the following fluctuations of the gauge field:
\bea
\delta A_x^j(t,r,x)&=&a_x^j(r)e^{i k_x x-i\omega t}\label{axkx},\\
\delta A_y^j(t,r,x)&=&a_y^j(r)e^{i k_x x-i\omega t}\label{aykx},\\
\delta A_t^j(t,r,x)&=&a_t^j(r)e^{i k_x x-i\omega t}\label{atkx} \eea
where $j$ runs over the three $SU(2)$ index. As we shall see the
equations of motion for the $\delta A_y^j$ fluctuation decouples and
can be analyzed separately.

\subsubsection{Unbroken phase}

In the normal phase, where $w(r)=0$, the 6 equations of motion for
the fluctuations \eqref{axkx} and \eqref{atkx} decouple in 2
(background independent) equations for $a_t^3, a_x^3$ and 4
equations coupling $a_x^1, a_x^2, a_t^1, a_t^2$. %Since rotational
%symmetry is not broken yet, we don't need to worry about
%fluctuations in the $y$ direction.

The sector with color indices $1-2$ gives us the mode that will
drive the instability towards the superfluid phase. Following this
mode through the phase transition, it will become the Goldstone
mode of the broken phase.

On the other hand, the sector with color index $3$, gives the
diffusive mode of the normal phase with a dispersion relation
$\omega=-i k^2$ \cite{Miranda:2008vb}. Going further from the origin
in the $\omega$ complex plane, the quasinormal frequencies at zero
momentum are know analytically to be \cite{Miranda:2008vb}
$\omega=-i \frac32 n$, with $n$ being a positive integer number. The
fate of these modes in the broken phase will be address in the
following sections.

The behavior in the near horizon limit of the four equations of the
$1-2$ sector allows to fix two of the horizon parameters in term of
the others. In order to have a solution for this system we must use
two pure gauge solutions (see eq. \eqref{puregauge1}). Instead, for the $3$-sector
we need just one pure gauge solution
$a_t^3=-\omega,\, a_x^3=k$ in order to determine the system.

On the other hand, the 3 equations for the fluctuations in the $y$
component \eqref{aykx} are separated in one equation for $a_y^3$
(which is temperature independent) and two coupled equations for
$a_y^1, a_y^2$. In this case there is no reason to search for pure
gauge solutions because the system is well defined. The equation for $a_y^3$ was previously studied
in \cite{Miranda:2008vb} and was shown that there is no a solution satisfying the Dirichlet condition at
the AdS boundary compatible with the hydrodynamic approximation.
Then, these set of equations give two modes coming from the $1-2$ sector. One of them is going to give a diffusive mode in the broken phase and the remaining mode is going to be a pseudo-diffusive.

Note that at zero momentum there is no distinction between $x$ and $y$
sectors since the rotational symmetry is not broken yet.

\subsubsection{Broken phase for $a_x, a_t$ sector}

Now, with a non vanishing condensate we must solve the following system with $6$
equations and $6$ unknowns

\bea
 &&a_t^{3\,\prime\prime}+\frac{- w (2 i k_x\, a_t^2+2
\,a_x^1 \phi+i \omega\,  a_x^2)-2 r^3 f a_t^{3\,\prime}+a_t^3
\left(
   w^2+k_x^2\right)+k_x\, \omega\, a_x^3}{r^4 f}=0\,,\nn\\
&&a_t^{2\,\prime\prime}-\frac{-2 r^3 f a_t^{2\,\prime}+
a_t^2 w^2+k_x^2 a_t^2-i (2 k_x a_t^3 w+k_x a_x^1
\phi)+\omega
   a_x^3 w)+k_x \omega a_x^2}{r^4 f}=0\,,\nn\\
&&a_x^{3\,\prime\prime}-\frac{a_t^1 w \phi-\omega  (i\, a_t^2 w+k_x a_t^3+\omega a_x^3)-f\,r^3 a_x^{3\,\prime}
   \left(r f'+2 f\right)}{r^4 f^2}=0\,,\nn\\
&&a_x^{1\,\prime\prime}+\frac{k_x\, \omega \, a_t^1+\phi
(i k_x a_t^2+2\, a_t^3 w+2 i \omega  a_x^2)+r^3 f
a_x^{1\,\prime}
   \left(r f'+2 f\right)+a_x^1 \left(\phi^2+\omega ^2\right)}{r^4 f^2}=0\,,\nn\\
&&a_x^{2\,\prime\prime}+\frac{-i(\phi (k_x a_t^1+2 \omega
a_x^1)+\omega a_t^3 w)+k_x \omega a_t^2+r^3 f
   a_x^{2\,\prime} \left(r f'+2 f\right)+a_x^2 \left(\phi^2+\omega ^2\right)}{r^4 f^2}=0\,,\nn\\
&&a_t^{1\,\prime\prime}-\frac{k_x^2 a_t^1-2 r^3 f
a_t^{1\,\prime}+k_x\, \omega\, a_x^1+\phi (-
a_x^3 w+i k_x a_x^2)}{r^4
   f}=0\,, \label{axateoms}
\eea where prime means derivative with respect to the radial
coordinate $r$. The near horizon behavior of the equations allow us
to fix $3$ of the IR parameters. Then we need to use $3$ pure gauge
solutions in order to have a well posed system of equations. The
pure gauge solutions are parametrized by $\lambda_i$ and read \bea
\nn a_x^1=- \lambda_2 k_x\,, \quad a_x^2=i \lambda_1 w(r)\, , \quad
a_x^3 =-\lambda_1 k_x-i \lambda_3 w(r) \, , \\ a_t^1=\lambda_2
\omega +i \lambda_3 \phi(r)\, ,\quad a_t^2=-i \lambda_2 \phi(r) +
\lambda_3 \omega \,,\quad a_t^3= \lambda_1\omega
\,.\label{puregauge1} \eea
 We
found two kind of hydrodynamic modes in this sector: two sound modes
and a diffusive mode.

%\begin{table}[htb!]
%  \begin{center}
%\begin{tabular}{|c||c|c|c|}
% \hline
%    & $I$ & $II$ & $III$\\
%  \hline
%  $a_x^1$ & 0 & $-k_x$ & 0\\\hline
%  $a_x^2$ & $i w(r)$ & 0 & $-k_x$ \\\hline
%  $a_x^3$ & $-k_x$ & 0 & $-iw(r)$\\\hline
%  $a_t^1$ & 0 & $\omega$ & $i\phi(r)$\\\hline
%  $a_t^2$ & 0 & $-i\phi(r)$ & $\omega$\\\hline
%  $a_t^3$ & $\omega$ & 0 & 0\\\hline
%\end{tabular}
%\end{center}
%\caption{Pure gauge solutions used to solve the equations of motion for the fluctuations shown in the first column.}\label{puregauge}
%\end{table}

~

\noindent {\bf{Sound Modes:}} We have two sound modes that satisfy
the dispersion relation \be \omega(k_x)=\pm v_x k_x -i\Gamma_x
k_x^2\,.\label{disprel} \ee In a superfluid the ordinary sound is
due to fluctuations of the density. The hydrodynamic equations for
the two fluid model predicts the existence of a different sound
produced by temperature or entropy fluctuations. This sound is
called second sound and it depends strongly on the temperature.
Along this work the second sound velocity is going to be denoted
with $v_i$, with $i$ being the spatial direction index. The function
$\Gamma_x(T)$ gives the imaginary part of the dispersion relations
\eqref{disprel}. It is called second sound attenuation, and is
related with the mean free path of the quasiparticles.

\begin{figure}[t!]
\centering
\includegraphics[width=230pt]{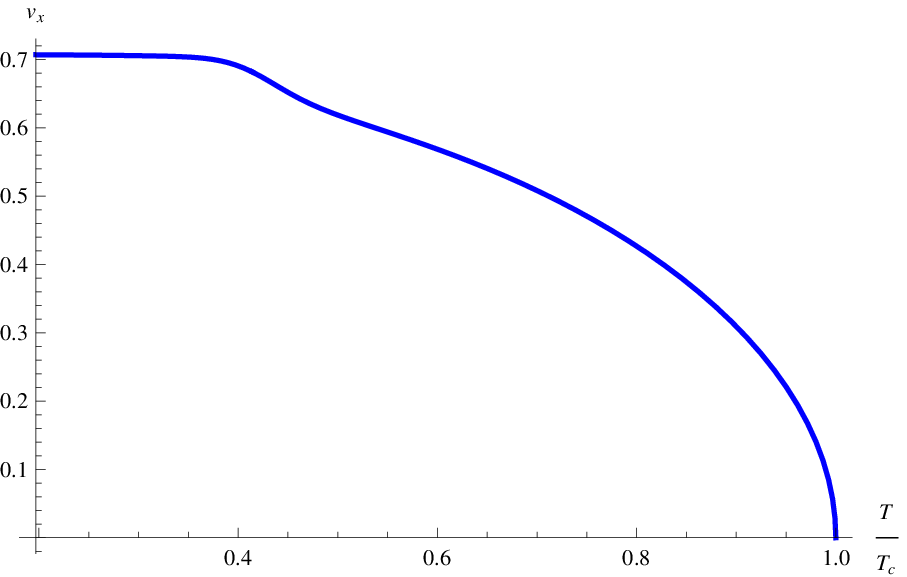}\hfill
\includegraphics[width=230pt]{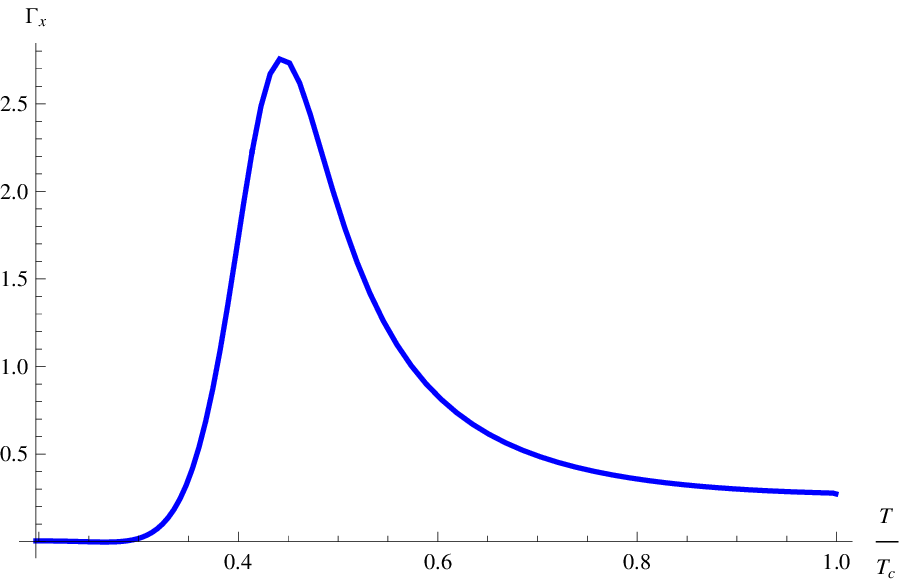}
\caption{Second sound velocity (left) and attenuation coefficient
(right) as a function of the temperature. The sound velocity
vanishes (as expected) at the critical temperature.} \label{secondS}
\label{Gamma}\end{figure}

In Figure \ref{secondS} we plot the second sound velocity as
function of the temperature. Since we have a second order phase
transition we expect everything to match at one and the other side
of the critical point. The notion of second sound only makes sense
in the two component fluid model of superconductivity, then is
expected the vanishing sound velocity at $T=T_c$ where just normal
fluid remains and there is no superfluid component. Furthermore, in
the normal phase this QNM becomes massive.

Another interesting feature of Figure \ref{secondS} is the change in
the slope for $T/T_c\sim0.45$. This is similar to the results
obtained in experiments with $^4He$ in \cite{Atkins, Peshkov} and
theoretically using a variational approach in \cite{Pitaevskii}.

In the right panel of Figure \ref{Gamma} we can see the temperature dependence of the
second sound attenuation. Note that there is a non vanishing
attenuation at the critical temperature ($\Gamma_s=0.273131 T_c$ for
$T=0.9992 T_c$) and it vanishes for very low temperatures. Similar
behavior for the attenuation near the critical temperature was
obtained in \cite{Buchel} for the normal sound of an $\mathcal{N}=2$
plasma and in \cite{Amado} for the second sound in the gravity dual
of an $s-$wave superconductor.

 ~

\noindent{\bf{Diffusive modes}}: These modes satisfy
the following relation
\be \omega(k_x)=-i D_x k^2_x%-i\gamma_x\,
,\label{difussive}
\ee
with $D_x$ the diffusive constant.
%and $\gamma_x$ a real parameter that shifts the pole from its unbroken phase position.Then, at zero momentum we have anon-vanishing $\omega(k_x)$.
The diffusion modes are expected in a
two fluid model of superfluidity because they are naturally related
with the normal fluid component. As was mentioned before in the
discussion of the unbroken phase this diffusive mode comes from the
$a_x^3, a_t^3$ sector of the equations of motion.

\begin{figure}[htb!]
\centering
\includegraphics[width=230pt]{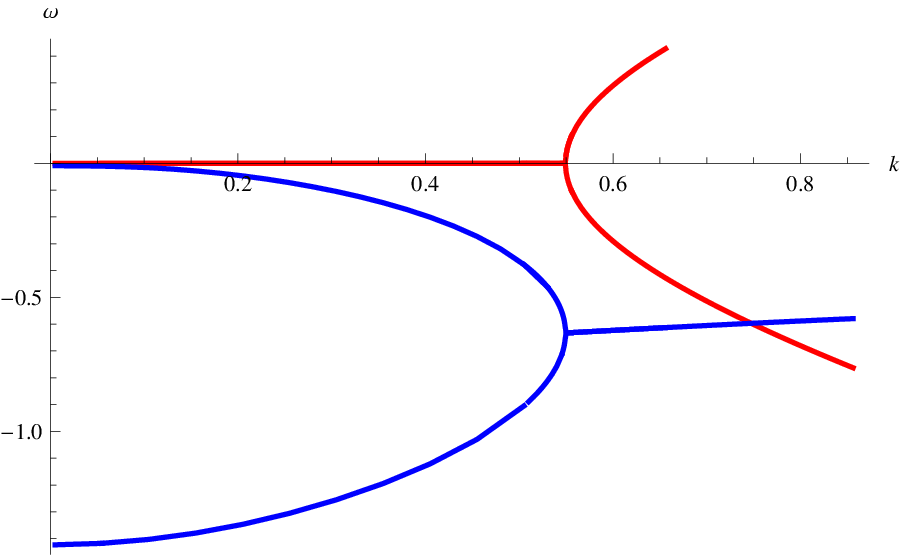}\hfill
\includegraphics[width=230pt]{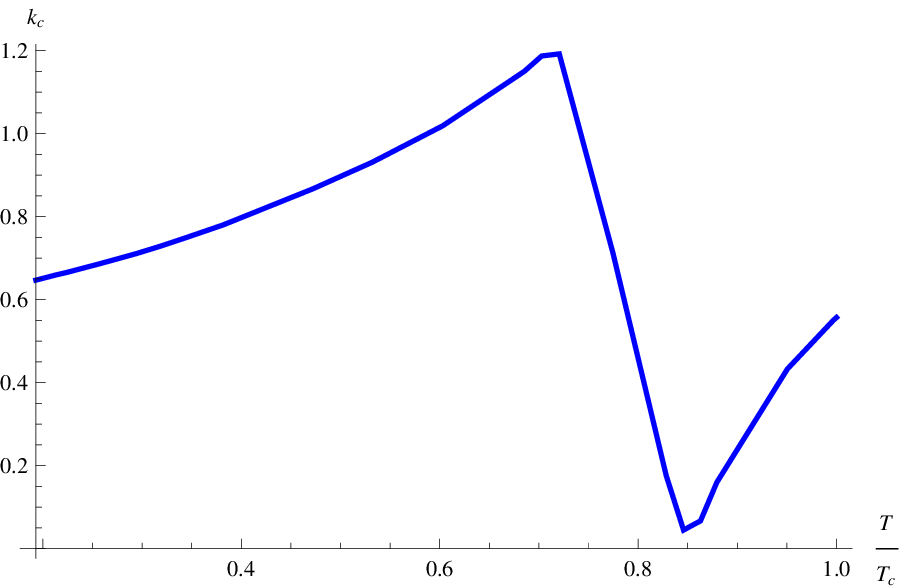}
\caption{(Left) Spectrum for the diffusive mode. The blue curves
correspond to the imaginary part of $\omega(k)$ while the red curves
to its real part. The lowest lying of the blue curves corresponds to
the first excited mode. We observe that at some value of $k=k_c$ it
meets the hydrodynamic mode and they acquire a non-vanishing real
part. (Right) Value of the momentum for which the diffusive and
first excited mode meets as a function of the temperature.}
\label{DiffX}\label{kcvst1}\end{figure}

Figure \ref{DiffX} shows the dispersion relation for the diffusive
modes at $T=0.77458 T_c$. The red lines are the real part of
$\omega$ and the blue lines its imaginary part. The branch with the
lowest imaginary part corresponds to the first excited mode. Note
that at some critical value of $k=k_c$ it meets the diffusive mode
and they both acquire a real part. This critical value of $k$ is
temperature dependent and its non vanishing at $T=T_c$. This is
because the blue ``bubble" is made of the hydrodynamic mode and the
first excited mode. We will see that $k_c(T)$ has a different
behavior for the diffusive modes in the transverse fluctuations.
Interestingly in \cite{Lippert, Davison, Jokela:2012vn,
DiNunno:2014bxa} this kind of ``bubble" behavior for the
hydrodynamic diffusive mode was observed for geometries originated
from different brane configurations.

It is interesting to discuss the existence of this $k_c$ considering
the discussion in \cite{Gauntlett,Amado:2013xya}. Since we expect
these modes to rule the late time behavior of a quench, we can
interpret this result as follows: during an inhomogeneous quench
with a characteristic wavelength $k$, the order parameter will have
an oscillatory or purely decaying behavior depending on weather $k$
is larger or smaller than $k_c$ at the final temperature of the
quench.

\subsubsection{Broken phase for $a_y$ sector}

The equations of motion for the the $a_y^i$ fluctuations decouple
from the $a_x^i$ perturbations previously analyzed and reads

\bea &&a_y^{3\,\prime\prime}-\frac{a_y^3\left(f\left(
w^2+k_x^2\right)-\omega^2\right)-f\left(r^3
a_y^{3\,\prime}\left(r\,f'+2f\right)-2\,i\, k_x
a_y^2w\right)}
{r^4f^2}=0\,,\nn\\
&& a_y^{1\,\prime\prime}+\frac{r^3 f a_y^{1\,\prime} \left(r f'+2
f\right)+a_y^1 \left(\phi^2-f\,k_x^2+\omega^2\right)+2 i
\omega
   a_y^2 \phi}{r^4 f^2}=0\,,\nn\\
&&a_y^{2\,\prime\prime}+\frac{-2 i \omega
a_y^1\phi-f\,r^3a_y^{2\,\prime} \left(r f'+2 f\right)+a_y^2 \left(f
\left(
   w^2+k_x^2\right)-\phi^2-\omega^2\right)-2 i k_x a_y^3 f w}{r^4 f^2}=0\,.\label{ayeoms}
\eea

The lowest lying solutions are a pseudo-diffusive mode of the form
\be
\omega(k_x)=-i \tilde{D}_x k^2_x-i\gamma(T)\,
\ee
%like the one written in equation \eqref{difussive}
and a proper diffusive mode with $\gamma=0$. Here ${\tilde D}_x$ is the diffusive constant
and $\gamma$ is a real parameter that shifts the pole from its unbroken phase position.
Then, at zero momentum we have a non-vanishing $\omega(k_x)$.

\begin{figure}[t!]
\centering
\includegraphics[width=230pt]{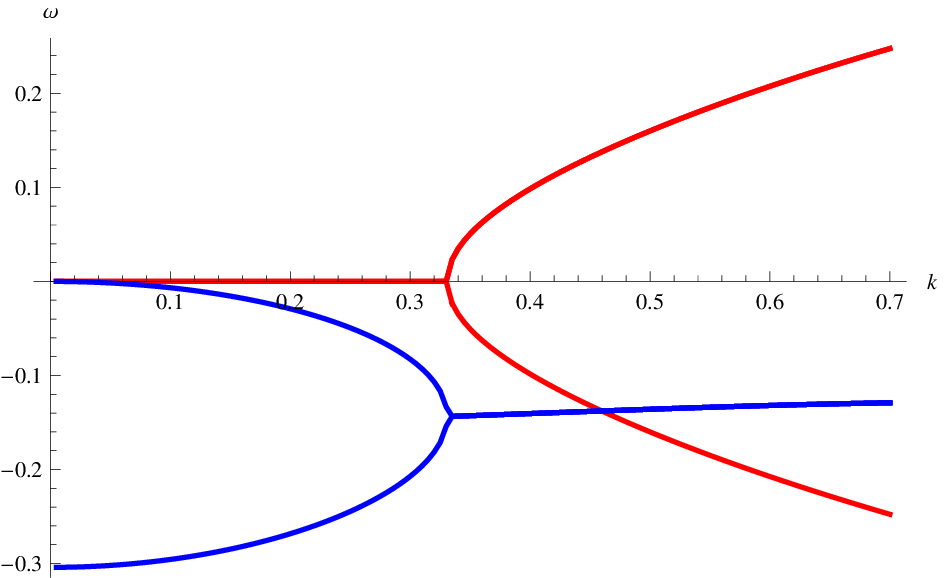}\hfill
\includegraphics[width=230pt]{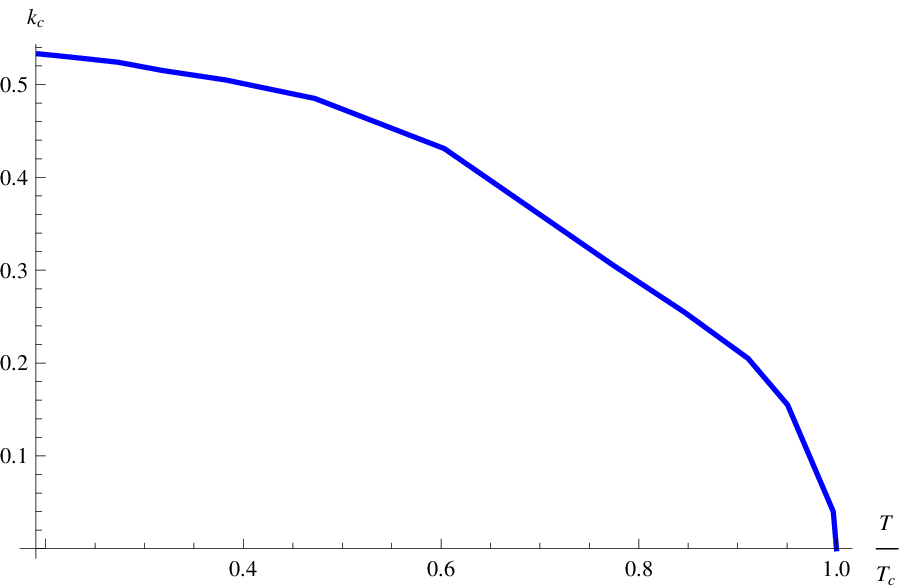}
\caption{(Left) Diffusive and pseudo-diffusive mode for the $a_y$
fluctuations. The blue and red curves correspond to the imaginary
and real part of the dispersion relation. (Right) Dependence on the
temperature of the critical value $k_c$ where the two poles meet and
acquire a non-vanishing real part.}
\label{kxaybubble}\label{kcxay}\end{figure}

A typical solution is shown in Figure \ref{kxaybubble} for $T=0.7745
T_c$ but its qualitative behavior is independent of the temperature.
The blue curves correspond to the imaginary part of $\omega(k)$
while the red ones correspond to the real part of $\omega(k)$. The
diffusive mode comes from the $a_y^3$ fluctuation in the unbroken
phase and the coupled equations for $a_y^{1,2}$ gives the lower blue
branch of the plot. Note the similitude with the diffusive mode for
the longitudinal fluctuation shown in figure \ref{DiffX}. This
similitude is just apparent because in the present case the two blue
branches correspond to a diffusive and a pseudo-diffusive mode. In
Figure \ref{DiffX} instead, the branches correspond with a diffusive
mode and the firs exited mode. Again we have a critical value for
the momentum $k$ where the modes meet and acquire non zero real
part. Figure \ref{kcxay} shows how this critical value $k_c$ depends
on the temperature.

The fact that the diffusive mode acquires a mass in the broken phase
was previously observed in \cite{Amado} in the context of
holographic $s-$wave superfluids. On the other hand, the coupling
between different (pseudo)diffusive modes, allows them to acquire a
real part in their mass or dispersion relation. A similar behavior
was observed in \cite{Amado:2013xya} in the context of $s-$wave
$U(2)$ superfluids.

%Interestingly in \cite{Lippert, Davison, Jokela:2012vn,
%DiNunno:2014bxa} this kind of "bubble" behavior for the hydrodynamic
%diffusive mode was observed for geometries originated from different
%brane configurations.

\subsection{Transverse modes}

In this section we are going to study the fluctuations that
propagate in the transversal direction with respect to the
condensate, i.e.
 those given by equation \eqref{transversal}. Due to the rotational symmetry of the unbroken phase
 we expect the same number of hydrodynamic modes that in the longitudinal case.

In this case we have two sets of fluctuations that can be
consistently solved. The first one (which we are going to call
sector $I$) reads

\bea
\delta A_x^j(t,r,y)&=&a_x^j(r)e^{i k_y y-i\omega t}\,,\\
\delta A_y^3(t,r,y)&=&a_y^3(r)e^{i k_y y-i\omega t}\,,\\
\delta A_t^3(t,r,y)&=&a_t^3(r)e^{i k_y y-i\omega t}\,, \eea with
$j=1,2$. From this sector we will find two sound modes and one diffusive mode.

The second consistent fluctuations (called sector $II$ from now on) read \bea
\delta A_x^3(t,r,y)&=&a_x^3(r)e^{i k_y y-i\omega t}\,,\\
\delta A_y^j(t,r,y)&=&a_y^j(r)e^{i k_y y-i\omega t}\,,\\
\delta A_t^j(t,r,y)&=&a_t^j(r)e^{i k_y y-i\omega
t}\,,\label{transdiffusive} \eea with $j=1,2$ and they lead to one diffusive and one
pseudo-diffusive modes.

\subsubsection{Unbroken phase}

Since we defined our transverse or parallel modes according to the
direction of the momentum with respect to the condensate, the
transverse and parallel modes coincide in the normal phase.
Nevertheless we will review them again here and discuss them
according on how they couple in the broken phase. This is important
since the continuity of the QNM across the phase transition will be
one of our checks to the results obtained numerically.

For sector $I$ we have two sets of equations. One of them couples
the fluctuations $a_t^3, a_y^3$ and are temperature independent. In
\cite{Miranda:2008vb} was shown that they give a diffusive mode. The
remaining two equations couples $a_x^1, a_x^2$ and we found that
they give the mode that drives the instability and evolves into the
Goldstone mode when $T<T_c$. This Goldstone mode is of course
massless and will give us at fine momentum a sound velocity. Thats
why we will call it also sound mode.

For the sector $II$ we have one decoupled equation for the $a_x^3$
fluctuation and four coupled equations for the remaining
fluctuations. The decoupled equation was previously studied in
\cite{Miranda:2008vb} and was shown that it does not give a solution
satisfying the boundary conditions compatible with the hydrodynamic
limit, but just massive modes. The remaining equations are going to give two
modes in the broken phase, one diffusive and one pseudo-diffusive mode.

\subsubsection{Broken phase - sector I}

The equations of motion related to this sector read

\bea
&&a_y^{3\,\prime\prime}-\frac{-k_y \omega a_t^3-f \left(r^3 a_y^{3\,\prime} \left(r f'+2 f\right)+i k a_x^2 w\right)+
a_y^3\left(f w^2-\omega ^2\right)}{r^4 f^2}=0\,,\nn\\
&&a_t^{3\,\prime\prime}-\frac{-2 r^3 f a_t^{3\,\prime}+a_t^3
\left(w^2+k_y^2\right)+w (2 g\, a_x^1 \phi+i
\omega a_x^2)+k_y
   \omega  a_y^3}{ r^4 f}=0\,,\nn\\
&&a_x^{1\,\prime\prime}+\frac{2\phi(a_t^3 w+i
\omega  a_x^2)+r^3 f a_x^{1\,\prime} \left(r f'+2 f\right)+a_x^1
   \left(\phi^2-f\,k_y^2+\omega ^2\right)}{r^4 f^2}=0\,,\nn\\
&&a_x^{2\,\prime\prime}-\frac{-a_x^2 \left(
\phi^2-f\,k_y^2+\omega ^2\right)+i \left(\omega\,  a_t^3
w+2 \omega  a_x^1
   \phi+i r^3 f a_x^{2\,\prime} \left(r f'+2 f\right)+k_y \,a_y^3 f\, w\right)}{r^4 f^2}=0\,.\label{sector1eoms}
\eea

We again need to find a pure gauge solution \bea a_x^1=0\,,\quad
a_x^2=-i w(r)\,,\quad a_y^3=-k_y\,,\quad a_t^3=\omega \eea in order
to use the determinant method.
 Then, we find from
\eqref{sector1eoms} two sound modes and one diffusive mode.

%\begin{table}[htb!]
%  \begin{center}
%\begin{tabular}{|c||c|}
% \hline
%    & $I$ \\
%  \hline
%  $a_x^1$ &$ 0$ \\\hline
%  $a_x^2$ &$ -iw(r)$ \\\hline
%  $a_y^3$ & $-k_y$ \\\hline
%  $a_t^3$ & $\omega$ \\\hline
%\end{tabular}
%\end{center}
%\caption{Pure gauge solution used to solve the equations of motion for the fluctuations shown in the first column.}\label{puregaugecaseI}
%\end{table}

~

\noindent {\bf{Sound modes:}} We have two sound modes that satisfy
the dispersion relation \be \omega(k_y)=\pm v_s k_y -i\Gamma_y
k_y^2\,,\label{disprely} \ee in the hydrodynamic limit

\begin{figure}[t!]
\centering
\includegraphics[width=230pt]{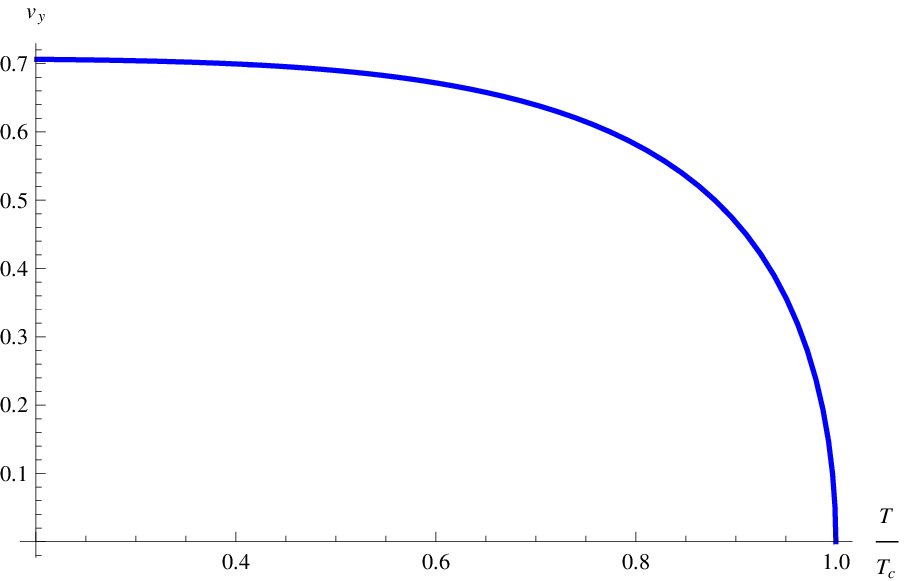}\hfill
\includegraphics[width=230pt]{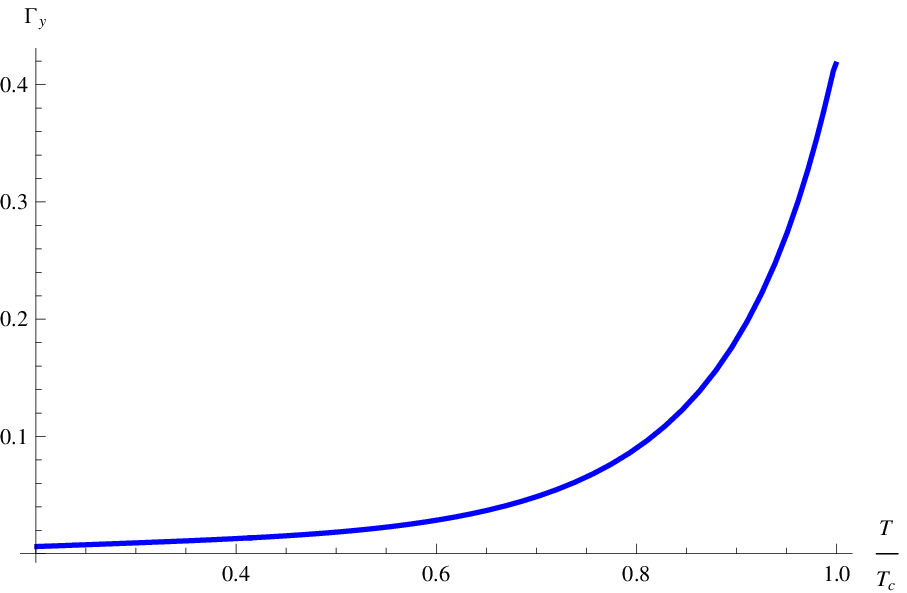}
\caption{Second sound velocity (left) and attenuation coefficient
(right) in the transversal direction as a function of the
temperature. The sound velocity vanishes (as expected) at the
critical temperature.} \label{DiffY}\label{secondSY}\end{figure}

In figure \ref{secondSY} we plot the second sound velocity as a
function of the temperature. Again, we observe the expected
vanishing behavior at $T=T_c$. On the other hand, in the low
temperature limit we see that it has the same value that in the
longitudinal case. In Figure \ref{secondSY} we observe the
transverse second sound attenuation as a function of the
temperature. Note that as in the longitudinal case it tend to zero
in the low temperature limit and takes a finite value at the
critical temperature ($\Gamma_y=0.4175T_c$ at $T=0.9992T_c$). This
behavior is analog to what was found in \cite{Amado} for the
$s-$wave case.

~

~

\noindent {\bf{Diffusive mode:}} The low momentum dispersion
relation for a diffusive mode reads \be
\omega(k_y)=-i D_y k^2_y%-i \gamma_y(T),
\ee and the qualitative behavior is analog to that in Figure
\ref{DiffX} where the lowest blue line is the imaginary part of the
first excited mode and the upper blue line is the pseudo-diffusive
mode. Again, they meet at some critical value $k_c$ and its
qualitative behavior is similar to the one shown in Figure
\ref{kcvst1}.

\subsubsection{Broken phase - sector  II}

The equations of motion for this sector read

\bea &&a_x^{3\,\prime\prime}+\frac{-a_t^1 w\, \phi+i
 \omega a_t^2 w+r^4 f a_x^{3\,\prime} f'+2 r^3 f^2
   a_x^{3\,\prime}+a_x^3 \left(\omega ^2-k_y^2 f\right)+i k_y a_y^2 f\, w}{r^4 f^2}=0\,,\nn\\
&&a_y^{1\,\prime\prime}+\frac{k_y \omega  a_t^1+i  \phi
(k_y a_t^2+2 \omega  a_y^2)+r^3 f\, a_y^{1\,\prime} \left(r f'+2
   f\right)+a_y^1\left(\phi^2+\omega ^2\right)}{r^4 f^2}=0\,,\nn\\
&&a_y^{2\,\prime\prime}+\frac{-i \phi(k_y a_t^1+2 \omega
a_y^1)+k_y \omega a_t^2-i  k_y a_x^3f\, w+r^3 f
   a_y^{2\,\prime} \left(r f'+2 f\right)+a_y^2 \left(\left(\phi^2-f\, w^2\right)+\omega ^2\right)}{r^4 f^2}=0\,,\nn\\
&&a_t^{1\,\prime\prime}-\frac{-2 r^3 f\, a_t^{1\,\prime}+k_y^2
a_t^1+\phi(-a_x^3 w+i k_y a_y^2)+k_y \omega
a_y^1}{r^4
   f}=0\,,\nn\\
&&a_t^{2\,\prime\prime}-\frac{-2 r^3 f a_t^{2\,\prime}+a_t^2
\left( w^2+k_y^2\right)-i (\omega a_x^3 w+k_y
a_y^1 \phi)+k_y
   \omega  a_y^2}{r^4 f}=0\,.\label{sector2eoms}
\eea

Solving this equations of motions and using the pure gauge solutions
\bea a_x^3=0\,,\quad a_y^1=-\lambda_1 k_y\,,\quad a_y^2 = -\lambda_2
k_y \,,\quad a_t^1= \lambda_1 \omega + i \lambda_2 \phi(r)\,,\quad
a_t^2= -i \lambda_1 \phi(r)+\lambda_2 \omega\,,  \eea we find a
pseudo-diffusive and a diffusive mode in the hydrodynamic regime.

%\begin{table}[htb!]
%  \begin{center}
%\begin{tabular}{|c||c|c|}
% \hline
%    & $I$ & $II$ \\
%  \hline
%  $a_x^3$ & 0 & 0 \\\hline
%  $a_y^1$ & $-k_y$ & 0 \\\hline
%  $a_y^2$ & 0 & $-k_y$  \\\hline
%  $a_t^1$ & $\omega$ & $i\phi(r)$ \\\hline
%  $a_t^2$ &  $-i\phi(r)$ & $\omega$ \\\hline
%\end{tabular}
%\end{center}
%\caption{Pure gauge solutions used to solve the equations of motion for the fluctuations shown in the first column.}\label{puregaugecaseII}
%\end{table}

~

\noindent{\bf{Pseudo-diffusive mode:}} as in the longitudinal case,
we have pseudo-diffusive and diffusive mode ($\gamma_y=0$ in the eq.
below) for the system obtained from the ansatz
\eqref{transdiffusive}. As before, the dispersion relation reads

\be
\omega(k_y)=-i D_y k^2_y-i \gamma_y(T).
\ee

\begin{figure}[htb!]
\centering
\includegraphics[width=230pt]{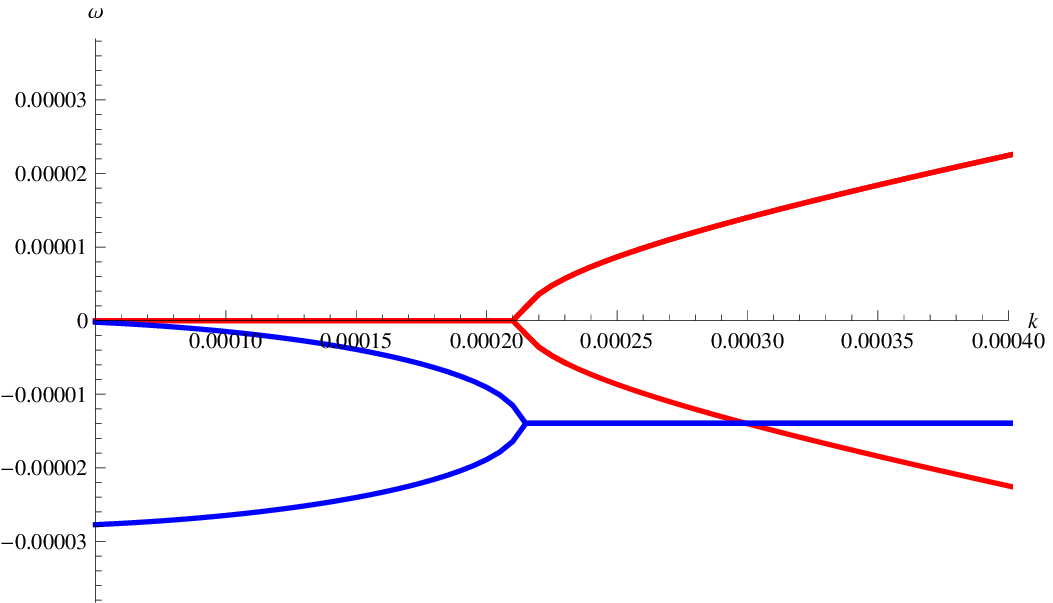}\hfill
\includegraphics[width=230pt]{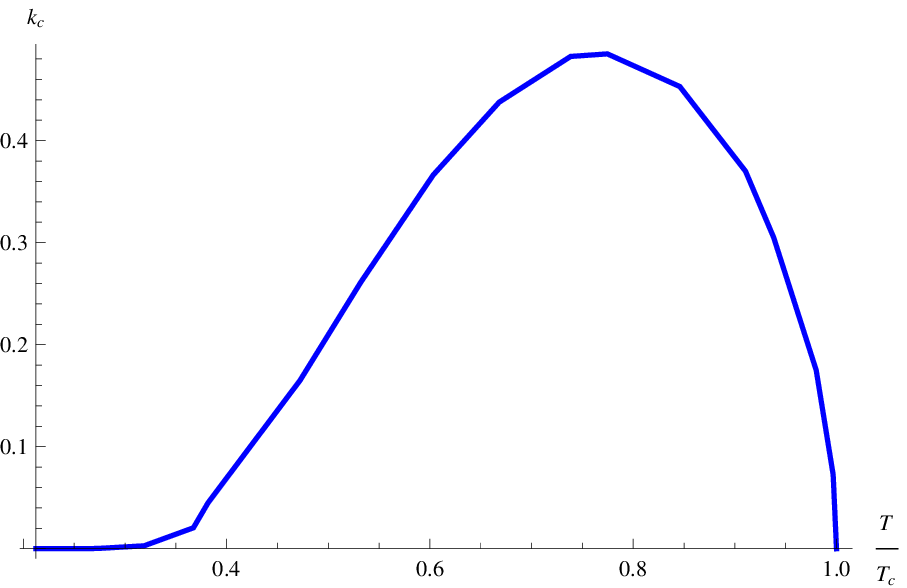}
\caption{(Left) (Pseudo-)diffusive modes for the transverse sector
$II$ fluctuations at $T=0.2732 T_c$. (Right) Critical value of $k$
as a function of the temperature.}
\label{Diffy}\label{kcyay}\end{figure}

In Figure \ref{Diffy} we plot these pseudo-diffusive modes of the
transverse fluctuations at $T=0.2732 T_c$. Again we have the
``bubble" behavior, i.e. the meeting of two modes at certain value
$k=k_c$ and the analysis of the spectrum is analog to that performed
for the $a_y$ fluctuations in the longitudinal mode. Figure
\ref{kcyay} also shows how this critical value $k_c$ depends on the
temperature. It has a maximum value at the temperature at which the
two diffusive modes are the furthest one from the other at zero
momentum. Moreover, at very low temperatures and near the critical
temperature they are almost at the same place on the imaginary axis.
This fact produces a small $k_c$ in those regimes.

~

~

\section{Conclusions}\label{concl}

In this paper we look for hydrodynamic modes of a $2+1$ dimensional
$p-$wave superfluid. In order to obtain this modes we review its
dual gravity background in the probe limit, which is an
asymptotically AdS planar charged black hole with a $SU(2)$ gauge
field living on it. We work in the non-backreacting limit and we use
holography to compute the poles of the current-current retarded
Green functions of the dual field theory. Note that in this
bottom-up approach we can not say anything about the microscopic
theory that originate the superconducting character of the material
but we can study its phenomenology. The AdS/CFT dictionary allow us
to obtain the poles of the retarded Green function through the
computation of the QNM of the gravity dual. Then with this plain in
mind we computed the QNM in the geometry dual to the $p-$wave
superconductor. We separate the fluctuations of the gauge field in
two sectors which we called longitudinal and transverse because they
propagate parallel or orthogonal with respect to the direction of
the condensate respectively. As expected by the rotational symmetry
present in the unbroken phase we have the same number of
hydrodynamic modes in both sectors. For the longitudinal modes we
find two subsystem of equations, the first one presents one
diffusive mode and two sound modes. We computed the velocity of the
second sound for these modes and its attenuation as a function of
the temperature. We noted that the diffusive mode meets an excited
mode at some value of the momentum $k_c$ and we made a plot of its
temperature dependence. The second subsystem leads to one
pseudo-diffusive and one diffusive modes. We observe that the
behavior of $k_c(T)$ for this modes is different that in the cases
where the diffusive modes meet the first excited mode. On the other
hand, we obtain the same kind of modes in the
 transverse sector. One diffusive mode and two sound modes for what was called sector $I$ and one pseudo-diffusive and one diffusive modes
 for sector $II$. Again, we plot the second sound velocity as a function of the temperature and observe that it has the same behavior that in
 the longitudinal case at low temperatures and at $T\sim T_c$.

In view of the work \cite{Gauntlett}, the modes found on the present
 paper can be viewed as responsible of the late time behavior of a $p-$wave
 superconductor. In particular we noticed a phase transition from a
 non oscillatory regime to an oscillatory one for inhomogeneous
 quenches with a characteristic wave number greater than $k_c$ and followed the
 behavior of $k_c$ as a function of the temperature.
 As a future work it would be interesting to explicitly check this affirmation by quenching the $p-$wave and computing its late time behaviour.
 Moreover it could be interesting to see if the hydrodynamic modes that we found could also be obtained using the hydrodynamic equations of \cite{Hoyos}.

Another possible extension to this work would be to analize the quasinormal modes of the so called $s+p-$wave phase introduced in \cite{Amado:2013lia}.
This model arises naturally in the low temperature regime of \cite{Amado:2013xya} and field theory calculations made in \cite{Gusynin:2003yu} suggest that
roton-like exitations may be found.

\section*{Acknowledgements}

We would like to thank Daniel ``epic" Are\'an, David Blanco, Richard
Davison, Nikolaos Kaplis, Steffen Klug, Guille Silva and Jan Zaanen
for useful discussions and comments. We would also like to thank
ICTP's Spring School in String Theory and related topics where part
of this work was done. R.A. would like to thank to Lorentz Leiden
Institute for cordiality during his visit. I.S.L. would like to
thank Romi and el Fresco for hospitality during his visit to
Bariloche and ICTP where part of this work was done in the context
of the STEP program. The work of R.A. was supported by CONICET, CNEA
and Universidad Nacional de Cuyo, Argentina.

\end{document}